\documentclass[prd,twocolumn,showpacs,preprintnumbers,amsmath,nofootinbib,amssymb,floatfix]{revtex4-1}
\usepackage{graphicx}
\usepackage{epsfig,float}
\usepackage[sort&compress]{natbib}
\usepackage{subfigure}
\usepackage{amsmath}
\usepackage{amsfonts}
\usepackage{cancel}
\usepackage{amssymb}
\usepackage{hyperref}
\usepackage{color}
\begin{document}
\arraycolsep1.5pt

\newcommand{\Slash}[1]{\ooalign{\hfil/\hfil\crcr$#1$}}

\title{Role of a triangle singularity in the $\gamma p\rightarrow K^+ \Lambda(1405)$ reaction}

\author{En Wang}
\affiliation{Department of Physics, Zhengzhou University, Zhengzhou, Henan 450001, China}

\author{Ju-Jun Xie}
\affiliation{Institute of Modern Physics, Chinese Academy of
Sciences, Lanzhou 730000, China} 

\author{Wei-Hong Liang}
\affiliation{Department of Physics, Guangxi Normal University,
Guilin 541004, China}

\author{Feng-Kun Guo}
\affiliation{CAS Key Laboratory of Theoretical Physics, Institute of Theoretical Physics,
Chinese Academy of Sciences, Beijing 100190, China}

\author{Eulogio Oset}
\affiliation{Institute of Modern Physics, Chinese Academy of
Sciences, Lanzhou 730000, China \\ and Departamento de
F\'{\i}sica Te\'orica and IFIC, Centro Mixto Universidad de
Valencia-CSIC Institutos de Investigaci\'on de Paterna, Aptdo.
22085, 46071 Valencia, Spain}

\date{\today}

\begin{abstract}
We show the effects of a triangle singularity mechanism for the
$\gamma p \to K^+\Lambda(1405)$ reaction. The mechanism has a
$N^*$ resonance around 2030~MeV, which decays into $K^* \Sigma$. The
$K^*$ decays to $K^+ \pi$, and the $\pi \Sigma$ merge to form the
$\Lambda(1405)$. This mechanism produces a peak around $\sqrt{s} = 2110$~MeV,
and has its largest contribution around cos$\theta=0$. The addition
of this mechanism to other conventional ones,  leads to a good
reproduction of ${\rm d}\sigma/{\rm dcos}\theta$ and the integrated
cross section around this energy, providing a solution to a problem
encountered in previous theoretical models.

\end{abstract}

\maketitle
\section{Introduction}
\label{sec:introduction}

Triangle singularities were discussed by Landau \cite{Landau:1959fi} and affect physical processes driven by a Feynman diagram with three intermediate states. The triangle singularities appear in a particular situation, where all intermediate states are placed on shell, and the particles move along the same direction. Even then, the singularities appear in a situation in which a classical analogy can be established where an original particle A decays into two particles 1 and 2, particle 1 decays into an external particle B and an internal particle 3, and finally particles 2 and 3 merge into an external particle C. The classical situation requires that particle 3 moves along the same direction and faster than particle 2 to catch up for the delay in its production. This is the essence of Coleman-Norton theorem when applied to a triangle diagram~\cite{Coleman:1965xm}. A simple analytical formula to impose these conditions is given in Ref.~\cite{guobayar}.

Although the study of these singularities dates from long ago~\cite{Landau:1959fi}, it is only recently, with the observation of many decay modes of particles by BESIII, Belle, Babar, LHCb Collaborations, that clear examples of such singularities have emerged. One of these examples is given in the $\eta(1405)\to \pi a_0(980)$ and $\eta(1405)\to\pi f_0(980)$ decay in Ref.~\cite{BESIII:2012aa}, studied in Refs.~\cite{Wu:2011yx,Wu:2012pg,Aceti:2012dj}, where an abnormal isospin violation in the second reaction is observed which is traced to one such singularity in which particles 1, 2, and 3 are $K^*$, $\bar{K}$, $K$, respectively, with the external particle B being a pion and the $\bar{K}K$ fusing to give the external C particle, $a_0(980)$ or $f_0(980)$. Another successful example was recently given in the decay of the $a_1(1260)$ axial vector resonance into $K^*\bar{K}$, with $K^*\to K \pi$, and $\bar{K}K$ fusing to give the $f_0(980)$ state. It was suggested in Ref.~\cite{Liu:2015taa} and shown in Refs.~\cite{Ketzer:2015tqa,Aceti:2016yeb} that the triangle diagram corresponding to this process gave a peak around 1420~MeV, providing a natural interpretation of the experimental observation by the COMPASS collaboration \cite{Adolph:2015pws}, where this peak was associated with a new resonance, the ``$a_1(1420)$".

Another recent issue that has stirred some discussion is the possibility that the narrow peak seen in the $J/\psi p$ mass distribution by the LHCb Collaboration~\cite{Aaij:2015tga,Aaij:2015fea}, dubbed as $P_c(4450)$, could be induced by a triangle singularity where the $\Lambda_b$ decays into $\Lambda(1890)$ and $\chi_{c1}$, the $\Lambda(1890)$ decays to $K^- p$ and the $\chi_{c1}p\to J/\psi p$~\cite{Guo:2015umn,Liu:2015fea}. A similar mechanism~\cite{Guo:2016bkl} would also be responsible for the peak seen at the same energy in the $\Lambda_b\to J/\psi \pi^- p$ reaction~\cite{Aaij:2016ymb,Aaij:2014zoa}. The fact that the energy 4450~MeV corresponds to a triangle singularity and to the threshold of $\chi_{c1}p\to J/\psi p$ enhances the peak structure over other possible configurations. The issue has been recently discussed in Ref.~\cite{guobayar}, where it was shown that, should the quantum numbers of the peak correspond to $3/2^-$, $5/2^+$, as currently suggested by the experimental analysis, it would require $p$- and $d$-waves in $\chi_{c1}p$, respectively, weakening and broadening the peak structure to the point that it cannot account for the experimentally observed narrow peak. For other quantum numbers that require $\chi_{c1}p$ in an $s$-wave, the possibility that these singularities account for the experimental peak would not be ruled out.

In the present work, we show one example of triangle singularity
that naturally explains the peak around $\sqrt{s}=2110$~MeV of the
$\gamma p\to K^+\Lambda(1405)$ reaction observed in Ref.~\cite{moriya},
which has resisted a theoretical interpretation so far.

Before the $\gamma p\to K^+\Lambda(1405)$ experiment was performed, there was a prediction based on a basic contact mechanism in Ref.~\cite{nacher}. The experiment was performed nine years later at Spring8/Osaka in Ref.~\cite{niiyama} and extended in Refs.~\cite{moriya,Schumacher:2013vma}.

The data of Refs.~\cite{moriya,Schumacher:2013vma} were analyzed in Refs.~\cite{luis1,luis2,mai} without an explicit model for  the reaction, parametrizing the production of $K^+MB$ ($MB$ standing for $\pi\Sigma$ and $\bar{K}N$) and letting the $MB$ system interact in coupled channels, where the two $\Lambda(1405)$ states are generated. The works independently show the need for two $\Lambda(1405)$ states, one narrow around 1420~MeV and another one broad at around $1350\sim 1380$~MeV, as predicted by all different works on the chiral unitary approach~\cite{ollerulf,jidocal,carmen,hyodo,borawolfrom,boraulf,ikada}.

Detailed models for the reaction have also been done before the detailed results of Ref.~\cite{moriya}. In Ref.~\cite{cotanch}, estimates were done for the
reaction, emphasizing crossing symmetry and duality, and the
$\Lambda(1405)$ was produced via $u$-channel contribution{\bf s}. Very
small ($\sim 1$ nb) and isotropic cross sections are predicted in
the model.  In Ref.~\cite{choi}, along similar lines, the $K_1(1270)$ exchange is included,
producing angular distributions. An effective
Lagrangian model approach is done in Ref.~\cite{nam}, where the 
$t$-channel $K$ and $K^*$ exchanges are considered in addition to the  $s$-channel $N$ exchange that is found dominant. The cross sections
obtained are rather small compared to experiment, and there is no peak in
the integrated cross section around $\sqrt{s}=2110$~MeV. 

The most complete theoretical work is the one of Ref.~\cite{jido},
based on the chiral unitary approach, keeping the different diagrams
where the photon can couple with special regard for gauge
invariance. The work is in line with similar works done in related
reactions, $K^- p \to \gamma \Lambda(1405)$~\cite{tokinacher} and
kaon photo- and electroproduction on the proton~\cite{borasoy}. The
model considers the production of the $M_j B_j$ meson-baryon
channels in $\gamma p \to K^+ M_j B_j$ that couple to $\pi \Sigma$
 upon final state interaction, with $M_j B_j \equiv K^- p$, $\bar{K}^0 n$,
$\pi^0 \Lambda$, $\pi^0 \Sigma^0$, $\eta n$, $\eta \Sigma^0$, $\pi^+
\Sigma^-$, $\pi^- \Sigma^+$, $K^+ \Xi^-$, $K^0 \Xi^0$. The model
contains $K$- and $K^*$-exchanges and requires the introduction
of some phenomenological contact terms and the addition of form
factors, which are fitted to the data of Refs.~\cite{moriya,mori2}. A
fair reproduction of the data is obtained, but it is textually
quoted in the work that ``for the $\gamma p \to K^+ \pi^0 \Sigma^0$ (the
reaction that filters isospin $I=0$ in $\pi\Sigma$) at $W = 2.0$ GeV (with $W \equiv \sqrt{s}$), there
is a sharp rise in the data at ${\rm cos} \theta$ = 0, while a
rather smooth behavior is found in the calculated counterpart".

It is thus a common feature of all theoretical calculations that
they cannot get the peak seen in the integrated cross section for
$\gamma p \to K^+ \Lambda(1405)$ around $W = 2.1$ GeV, corresponding to $E_\gamma=1.9$ GeV in the laboratory frame (see Fig. 16
of Ref.~\cite{moriya}), which is associated with a large contribution
in the differential cross section around ${\rm cos} \theta \simeq
0$.

In the present work we bring a novel idea, with a mechanism not
considered before, which produces an enhancement of the cross
section around $W = 2.11$ GeV, peaking at ${\rm cos} \theta \simeq
0$. The
mechanism is tied to a triangle singularity that develops from a
vector-baryon ($VB$) state predicted in Refs.~\cite{angelsvec,angelselsa}, coupling
strongly to $K^* \Sigma$. This state was supported experimentally
from the $\gamma p \to K^0 \Sigma^+$ reaction close to the $K^*
\Lambda$ threshold studied in Ref.~\cite{ewald}. The mechanism then
proceeds as follows: The $VB$ state is formed from $\gamma p$, then
it decays into $K^* \Sigma$ and the $K^*$ decays into $K^+ \pi$,
finally the $\pi \Sigma$ merge into the $\Lambda(1405)$ state. This
triangle diagram has a singularity around $W = 2.11$ GeV and its
structure makes it contribute mostly around ${\rm cos}\theta =0$.
We shall show that the position, strength and width match with the
experimental data, and brings an unexpected solution to a persistent
problem encountered with the use of conventional models.

\section{Formalism}\label{sec:formalism}

\subsection{Baryon resonances from the vector-baryon interaction}

In Ref.~\cite{angelsvec} the interaction of vector mesons with the
octet of baryons was studied. The interaction was found attractive
in many cases, to the point that some baryonic states emerged as
molecular states owing to this interaction. One of the states
observed, which is the one of relevance to the present work, was a
state with $I =1/2$ and strangeness $S=0$ and spin-parity $J^P =
1/2^-$ or $3/2^-$ that peaks around $2000$~MeV, with a width around
100~MeV. The state couples to $\rho N$, $\omega N$, $\phi N$, $K^*
\Lambda$, and $K^* \Sigma$, but the largest coupling is to the latter channel. The state qualifies as basically a $K^* \Sigma$
molecular state, with $K^* \Lambda$ as the main decay channel. The
width obtained is a lower bound, because the state can also decay
into pseudoscalar-baryon. The mixture of vector-baryon and
pseudoscalar-baryon was undertaken later in Ref.~\cite{garzon} in
the light sector and in Ref.~\cite{lianguchino} in the charm sector.
The findings in Ref.~\cite{garzon} indicated that the states of mostly
vector-baryon nature were not much modified regarding their masses
by the mixture with pseudoscalar-baryons, but the widths become
bigger.

The predicted state got a boost from the study of the $\gamma p \to
K^0 \Sigma^+$ reaction close to the $K^* \Lambda$ and $K^* \Sigma$
thresholds, where the cross section shows a sudden drop and the
differential cross section experiences a transition from a
forward-peaked distribution to a flat one~\cite{ewald}. This
phenomenon was interpreted in Ref.~\cite{angelselsa} in terms of the $N^*$ state discussed
above in Ref.~\cite{angelsvec}, and as a result the mass and width
were determined with more precision. It was found that $M_{N^*} = 2030$~MeV. With this
increased mass there is more phase space for $K^* \Lambda$ decay and
the width, within the model of Ref.~\cite{angelsvec}, is $\Gamma = 127$~MeV. Yet, should one mix the vector-baryon state with
pseudoscalar-baryons, the width would become appreciably bigger.
This is why this state was associated in Ref.~\cite{angelsvec} to
the $N^*(2080)(3/2^-)$ and $N^*(2090)(1/2^-)$ of the former PDG (Particle Data Group)
tables~\cite{amsler}, which have a width between $180-450$ and
$100-400$~MeV, respectively. We shall take tentatively, $\Gamma =
300$~MeV but will discuss the effects of other choices, and we shall
refer to this state as the $N^*(2030)$ in what follows.

\subsection{The mechanism for the $\gamma p \to K^+ \Lambda(1405)$ reaction}

\begin{figure}[t]
\includegraphics[width=0.4\textwidth]{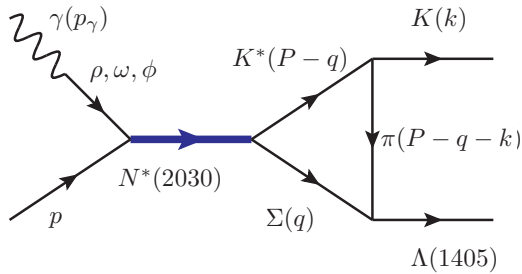}
\caption{Mechanism for the $\gamma p
\to K^+ \Lambda(1405)$ reaction involving the formation of
$N^*(2030)$ and its decay to $K^* \Sigma$.} \label{fig:feydiagram}
\end{figure}

The mechanism that we study is given in Fig.~\ref{fig:feydiagram}.
The photon is converted into a vector meson, $\rho^0$, $\omega$, and
$\phi$, according to the rules of the local hidden gauge
approach~\cite{hidden1,hidden2,hidden4} (see also practical rules in
Ref.~\cite{hideko}), which implements automatically the vector meson
dominance idea of Sakurai~\cite{sakurai}. The $\rho N$, $\omega N$,
and $\phi N$ are some of the coupled channels that generate the
$N^*(2030)$ and the couplings of this resonance to all the coupled
channels are evaluated in Ref.~\cite{angelsvec}. As mentioned
before, the largest coupling of the $N^*(2030)$ is to $K^*\Sigma$,
so, in Fig.~\ref{fig:feydiagram} the $N^*(2030)$ is allowed to decay
into $K^* \Sigma$. The $K^*$ will decay into $K^+ \pi$ and the $\pi
\Sigma$ can fuse to produce the $\Lambda(1405)$. This is the way we
can produce $K^+ \Lambda(1405)$ at the end. There would be nothing
special in this mechanism if it were just one more perturbative
diagram. However, it just happens that the diagram develops a
triangle singularity for a $\gamma p$ center-of-mass energy of
about $2110$~MeV, which renders this particular energy special and
the effects of the singularity show up clearly in the cross section
around this energy.

There is no need to evaluate the amplitude for the mechanism of
Fig.~\ref{fig:feydiagram} to know that there is a singularity at
that energy. For this, it is sufficient to apply the easy rule
obtained in Ref.~\cite{guobayar}，
\begin{eqnarray}
q^{\rm on}_+ = q^{a}_- \, , \label{eq:qonconstraint}
\end{eqnarray}
where $q^{\rm on}_+$ is the on-shell momentum of the $\Sigma$ in the center-of-mass frame of $\gamma p$ in the
$\gamma p \to K^* \Sigma$ transition,
\begin{eqnarray}
q^{\rm on}_+ =
\frac{\lambda^{1/2}(s,m^2_{K^*},m^2_{\Sigma})}{2\sqrt{s}},
\end{eqnarray}
and $q^{a}_-$ is the $\Sigma$ momentum for $\Sigma$ and $\pi$ in Fig.~\ref{fig:feydiagram} being on-shell simultaneously in the same frame in a special kinematical region (to be specified below),
\begin{eqnarray}
q^{a}_- = \gamma' (vE^*_2 - p^*_2),
\end{eqnarray}
with
\begin{eqnarray}
v &=& \frac{k}{E_{\Lambda^* (k)}}, ~~~~\gamma' =
\frac{1}{\sqrt{1-v^2}}
= \frac{E_{\Lambda^*}(k)}{m_{\Lambda^*}}, \\
E^*_2 &=& \frac{m^2_{\Lambda^*} + m^2_{\Sigma} -
m^2_{K^*}}{2m_{\Lambda^*}}, ~ p^*_2 =
\frac{\lambda^{1/2}(m^2_{\Lambda},m^2_{\pi},m^2_{\Sigma})}{2m_{\Lambda^*}},
\end{eqnarray}
where we define $\lambda(x,y,z)=x^2+y^2+z^2-2xy-2yz-2xz$.
One can see that $p^*_2$ and $E^*_2$ are the momentum and energy of the $\Sigma$ in the rest frame of the $\Lambda(1405)$
for the decay into $\pi \Sigma$, $v$ the velocity of the $\Lambda^*$
in the $\gamma p$ original rest frame, and $\gamma'$ the Lorentz
boost factor. The solution $q^{a}_-$ corresponds to a situation
where the $\Sigma$ in the rest frame of the $\Lambda^*$ goes in the
direction of the momentum of the $\Lambda^*$ in the $\gamma p$
center-of-mass frame. This makes the $\Sigma$ momentum smaller in
the $\gamma p$ center-of-mass frame and allows the $\pi$ emitted
from the $K^* \to K^+ \pi$ decay to catch up with the $\Sigma$, which
was emitted earlier, to form the $\Lambda^*$ in a classical picture,
according to the Coleman-Norton theorem~\cite{Coleman:1965xm}. Applying
Eq.~\eqref{eq:qonconstraint} it is easy to see that a singularity
appears  around $\sqrt{s} = 2110~{\rm~MeV}$, corresponding to placing on
shell simultaneously all three particles in the triangle diagram in the
case where $\vec{q}$ and $\vec{k}$ (see momentum of
Fig.~\ref{fig:feydiagram}) are parallel and go in opposite
directions. In other words the $\Sigma$ and the $\Lambda^*$ go in
the same direction in the $\gamma p$ center of mass frame. This is
called the parallel solution and is the only one giving rise to the
singularity.

The evaluation of the amplitude in Fig.~\ref{fig:feydiagram} is
straightforward and we follow the steps of Ref.~\cite{fcajorgi}. In
the $\int d^4q$ integral that we have with three propagators one performs
analytically the $q^0$ integration and the remaining $\int
d^3\vec{q}$ integration is done numerically.


\begin{figure}[t]
\includegraphics[width=0.2\textwidth]{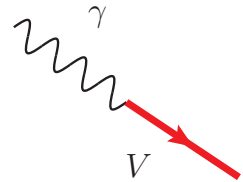}
\caption{Diagram for $\gamma$-$V$
conversion.} \label{fig:gammav}
\end{figure}

We need the Lagrangians~\cite{hidden1,hideko}
\begin{equation}
\mathcal{L}_{\gamma V} = -M^2_V\frac{e}{g}A_\mu <V^\mu Q>,
\label{eq:lagrangianV}
\end{equation}
where $M_V$ is the mass of the vector mesons, $g$ the coupling in
the local hidden gauge
\begin{eqnarray}
g=\frac{M_V}{2f_{\pi}}
\end{eqnarray}
with $f_{\pi}$ the pion decay constant ($f_{\pi} = 93$~MeV), and $e$
is the electric charge of the electron ($-|e|$), with $e^2/4\pi =
\alpha$, $<>$ is the trace of the SU(3) matrices, $V_{\mu}$ is the
ordinary SU(3) matrix for the vector mesons~\cite{hideko}, and $Q$ is
the diagonal matrix $Q={\rm diag}(2,-1,-1)/3$. The combination of
the diagram of Fig.~\ref{fig:gammav} gives rise to the amplitude
\begin{eqnarray}
-i\tilde{t} &=& i\frac{e}{g}\mathcal{C}_{\gamma V}
\epsilon_l(\gamma),
\end{eqnarray}
with $\epsilon_l(\gamma)$ the polarization vector of the photon
which replaces the vector polarization in the last vertex and has spatial
components, $l =1,~2$, since we work in the Coulomb gauge where
$\epsilon^0 = 0$ and $\vec{\epsilon} \cdot \vec{p_\gamma} = 0$,
where only the transverse photon polarizations are operative. The
coefficients $\mathcal{C}_{\gamma V}$ are given by
\begin{eqnarray}
\mathcal{C}_{\gamma V}\equiv  \left\lbrace\begin{array}{cc}
\frac{1}{\sqrt{2}}  &~~~~~~ \rho^0 \\
\frac{1}{3\sqrt{2}} &~~~~~~ \omega \\
-\frac{1}{3}        &~~~~~~ \phi   \end{array}\right. .
\label{eq:ampgV}
\end{eqnarray}


The $VB \to V'B'$ amplitude is given by
\begin{eqnarray}
-it_{VB,V'B'} &=&  -i \frac{g_{N^*VB}g_{N^*V'B'}}{\sqrt{s} - M_{N^*}
+ i \frac{\Gamma_{N^*}}{2}} \vec{\epsilon}(V) \cdot
\vec{\epsilon}(V'),\label{eq:ampgV}
\end{eqnarray}
and the couplings $g_{N^*VB}$ and $g_{N^*V'B'}$ are tabulated in
Ref.~\cite{angelsvec} and given in isospin basis by
\begin{eqnarray}
&& g_{N^*N\rho}=-0.3-0.5i,~~g_{N^*N\omega}=-1.1-0.4i, \nonumber
\\
&&g_{N^*N\phi}=1.5+0.6i,~~~~g_{N^*K^*\Sigma}=3.9+0.2i.
\end{eqnarray}

The $K^* \to K \pi$ vertex is also given by the local hidden gauge approach via the Lagrangian
\begin{equation}\label{eq:Lagrangian_VPP}
  \mathcal{L}_{VPP}=-ig \langle [P, \partial_{\mu}P] V^{\mu}\rangle
\end{equation}
with $P$ the SU(3) matrix for the pseudoscalar mesons (see Ref.~\cite{hideko}). Finally we need the coupling of the $\Lambda^*$ to $\pi \Sigma$ which is given in Ref.~\cite{jidocal} (we shall come back to this point).

\begin{figure}[h!]
\includegraphics[width=0.4\textwidth]{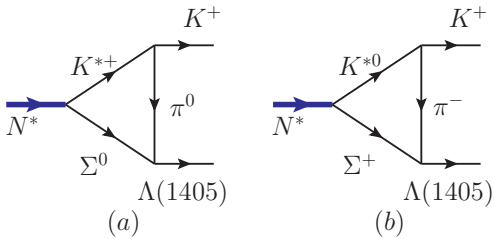}
\caption{Two charge configurations to be considered.} \label{fig:charge_cons}
\end{figure}

We also have to consider two charge configurations which are given in Fig.~\ref{fig:charge_cons}.

After proper isospin projections of the vertices, we finally can write the total amplitude as
\begin{eqnarray}
-it_T &=& \sum_m \sum_{i= \rho, \omega, \phi} i \frac{e}{g} C_{\gamma V_i} g_{N^* N V_i} \frac{i}{\sqrt{s}-M_{N^*}+i \frac{\Gamma_{N^*}}{2}}\nonumber \\
&&\times \left(\vec{\epsilon}(\gamma) \cdot \vec{\epsilon}_{K^*} \right) \int \frac{{\rm d}^4q}{(2\pi)^4} (-i) g^{(m)}_{N^*, K^* \Sigma}\nonumber \\
&& \times \frac{i}{(P-q)^2-m^2_{K^*}+i m_{K^*}\Gamma_{K^*}} \nonumber \\
&&\times ig \epsilon_j(K^*) (k-P+q+k)_j \mathcal{C}^{(m)} (-i) g^{(m)}_{\Lambda^* \pi \Sigma} \nonumber \\
&& \times \frac{i}{(P-q-k)^2-m^2_{\pi}+i \varepsilon} \frac{i2M_{\Sigma}}{q^2-M^2_{\Sigma}+i \varepsilon}, \label{eq:t_T}
\end{eqnarray}
with
\begin{eqnarray}
\mathcal{C}^{(m)}=\left\lbrace\begin{array}{cc}
\frac{1}{\sqrt{2}},  &~~~~~~ \pi^0 \\
1, &~~~~~~ \pi^-
\end{array}\right. ,
\label{eq:Cm}
\end{eqnarray}
\begin{eqnarray}
g^{(m)}_{\Lambda^* \pi \Sigma} &\equiv& g_{\Lambda^* \pi \Sigma} \left\lbrace\begin{array}{cc}
-\frac{1}{\sqrt{3}},  &~~~~~~ \pi^- \Sigma^+ \\
-\frac{1}{\sqrt{3}},  &~~~~~~ \pi^0 \Sigma^0 \\
\end{array}\right. , \nonumber \\
g^{(m)}_{N^*, K^* \Sigma} &\equiv& g_{N^*, K^* \Sigma} \left\lbrace\begin{array}{cc}
\sqrt{\frac{2}{3}},  &~~~~~~ K^{*0} \Sigma^+ \\
\frac{1}{\sqrt{3}},  &~~~~~~ K^{*+} \Sigma^0 \\
\end{array}\right. .
\label{eq:gm}
\end{eqnarray}

We also make the approximation, as in Refs.~\cite{angelsvec,fcajorgi} that the three-momenta of the vector mesons are small compared to their masses and hence
\begin{equation}
 \sum_{\rm pol} \epsilon_i (K^*) \epsilon_j (K^*) =\delta_{ij},
\end{equation}
which is valid since we are focusing on the energies close to the $K^*\Sigma$ threshold.

We also have an integral of $q_j$, with three propagators, where the only non-integrated vector is $\vec k$. Thus,
\begin{equation*}
  \int {\rm d}^3 q q_j \cdots =A k_j
\end{equation*}
from where $A=\int  {\rm d}^3q \frac{\vec k \cdot \vec q}{{\vec k}^2} \cdots$.

Taking this into account we obtain
\begin{eqnarray}
t_T &=& e  C_{\gamma N^*}~ C_T~ g^{(I=1/2)}_{N^*, K^* \Sigma}~g^{(I=0)}_{\Lambda^* \pi \Sigma} \frac{1}{\sqrt{s}-M_{N^*}+i \frac{\Gamma_{N^*}}{2}} \nonumber \\
&&\times \left[\vec{\epsilon}(\gamma) \cdot \vec k\right] ~i \int \frac{{\rm d}^4q}{(2\pi)^4} \frac{1}{(P-q)^2-m^2_{K^*}+i m_{K^*}\Gamma_{K^*}}  \nonumber \\
&& \times \frac{1}{(P-q-k)^2-m^2_{\pi}+i \varepsilon}
 \frac{2M_{\Sigma}}{q^2-M^2_{\Sigma}+i \varepsilon} \nonumber \\
&& \times \left( 2+\frac{\vec q \cdot \vec k}{\vec k^2} \right), \label{eq:t_T2}
\end{eqnarray}
where in $C_T$ we have combined the isospin factors and in $C_{\gamma N^*}$ the sum of contributions from $\rho$, $\omega$, $\phi$. Thus
\begin{equation}\label{eq:C_factors}
  C_T=-\frac{1}{\sqrt{2}},~~~~C_{\gamma N^*}=\sum_{i=\rho, \omega, \phi} C_{\gamma V_i} ~g_{N^* N V_i}.
\end{equation}

By performing the $q^0$ integration analytically we finally get an easy expression
\begin{eqnarray}
t_T &=& e ~ C_{\gamma N^*} ~C_T ~g^{(I=1/2)}_{N^*, K^* \Sigma}~g^{(I=0)}_{\Lambda^* \pi \Sigma}   \nonumber \\ & & \times \frac{1}{\sqrt{s}-M_{N^*}+i \frac{\Gamma_{N^*}}{2}}
\left[\vec{\epsilon}(\gamma) \cdot \vec k \right] ~\mathcal{I} \nonumber \\
&&\equiv B \times \left[\vec{\epsilon}(\gamma) \cdot \vec k\right],
\label{eq:t_T3}
\end{eqnarray}
which defines $B$, where $\mathcal{I}$ is the three-dimensional integral \cite{fcajorgi}
\begin{eqnarray}
\mathcal{I} &=&2M_\Sigma \int\frac{{\rm d}^3q}{(2\pi)^3} \frac{1}{8E_\Sigma\omega_{K^*}\omega_\pi} \frac{1}{k^0-\omega_\pi-\omega_{K^*}+i\frac{\Gamma_{K^*}}{2}} \nonumber \\
&&\times \frac{1}{P^0-\omega_{K^*}-E_\Sigma+i\frac{\Gamma_{K^*}}{2}}
 \,\frac{1}{P^0+E_\Sigma+\omega_\pi-k^0} \nonumber \\
&&\times \frac{1}{P^0-E_\Sigma-\omega_\pi-k^0+i\epsilon} \left(2+\frac{\vec{q}\cdot \vec{k}}{|\vec{k}|^2}\right)\nonumber \\
&& \times  \left\lbrace2P^0E_\Sigma+2k^0\omega_\pi - 2\left(E_\Sigma+\omega_\pi\right)\left(E_\Sigma+\omega_\pi+\omega_{K^*}\right)\right\rbrace, \nonumber \\
\end{eqnarray}
with $\omega_{K^*}=\sqrt{m^2_{K^*}+ \vec{q}^2}$, $E_\Sigma =\sqrt{M^2_\Sigma+\vec{q}^2}$ and $\omega_{\pi}=\sqrt{m^2_{\pi}+ (\vec{q}+\vec{k})^2}$.
Although formally convergent, the integral of $\mathcal{I}$ is regularized by a cut-off $\theta(\Lambda - |\vec q|)$ which appears in the chiral unitary approach \cite{angels} with $\Lambda =630 {~\rm~MeV}$.

The differential cross section is given by
\begin{equation}\label{eq:dCrossSection}
\frac{{\rm d}\sigma}{{\rm d}\Omega} =\frac{1}{64\pi^2 }\frac{2M_p~ 2M_{\Lambda^*}}{s} \frac{|\vec{k}|}{p_\gamma}\overline{\sum}\sum|t|^2.
\end{equation}

Since we average over the photon polarizations, we have
\begin{equation}\label{eq:polarization}
 \sum_{\rm pol} \epsilon_i(\gamma) \epsilon_j(\gamma) =\delta_{ij} -\frac{p_{\gamma_i} ~p_{\gamma_j}}{\vec p_{\gamma}^2},
\end{equation}
and with the term obtained in Eq. (\ref{eq:t_T3}) we have
\begin{equation}\label{eq:t_square}
\overline{\sum}\sum|t_T|^2 =\frac{1}{2} |B|^2 \left[ {\vec k}^2 -\frac{( \vec{k} \cdot \vec{p}_\gamma)^2}{\vec{p}_\gamma^2} \right]
\end{equation}
which goes as $(\sin \theta)^2$, with $\theta$ the angle between the kaon and the photon. The mechanism that we have produces a peak around $\sqrt{s} =2110$ ~MeV, peaking when $\cos \theta =0$, as a consequence of the photon being transversely polarized, a most welcome feature to solve the two problems encountered in the interpretation of the data.

\subsection{The $g_{\Lambda^* \pi \Sigma}$ coupling with two $\Lambda^*$ resonances}
Should there be just one $\Lambda^*$ resonance, we would take the coupling $g_{\Lambda^* \pi \Sigma}$ in the formula of Eq. (\ref{eq:t_T3}), but there are two $\Lambda (1405)$ states (see note in the PDG to the respect \cite{hyodoulf}).
To take this into account, we note that the $\Lambda (1405)$ is observed in the $\pi \Sigma$ channel. Hence, when we say that we produce the $\Lambda(1405)$ it actually means that we observe the production of $\pi \Sigma$ and integrate over the phase space of this state. Thus, what enters this evaluation is the $\pi \Sigma \to \pi \Sigma$ amplitude and the integration over the final $\pi \Sigma$ phase space. This amplitude gets the coherent sum of the two $\Lambda (1405)$ states. Should there be just one resonance we would have
\begin{equation}\label{eq:t_pisigma}
 t_{\pi \Sigma, \pi \Sigma} = \frac{g^2_{\Lambda^* \pi \Sigma}}{\sqrt{s}-M_{\Lambda^*}+i \Gamma_{\Lambda^*}/2}.
\end{equation}
The coherent sum of the two resonances still has a clear peak and can be roughly approximated by
\begin{equation}\label{eq:t_pisigma2}
 t_{\pi \Sigma, \pi \Sigma} = \frac{\tilde{g}^2_{\Lambda^* \pi \Sigma}}{\sqrt{s}-\tilde{M}_{\Lambda^*}+i \tilde{\Gamma}_{\Lambda^*}/2}.
\end{equation}
At the peak of the distribution we have
\begin{equation}\label{eq:t_pisigma3}
 \left|t_{\pi \Sigma, \pi \Sigma}\right|^2 = \frac{\left|\tilde{g}_{\Lambda^* \pi \Sigma}\right|^4}{\left( \tilde{\Gamma}_{\Lambda^*}/2\right)^2},
\end{equation}
and by using the input from Ref.~\cite{luis1} we obtain
\begin{equation}\label{eq:input}
  \tilde{\Gamma}_{\Lambda^*}= 65~ {\rm~MeV}, ~~~~~~~~~~~~~~~~~|\tilde{g}_{\Lambda^* \pi \Sigma}|^2=3.25.
\end{equation}
Since this amplitude is dominated by the first pole around 1380~MeV, we take this modulus and the phase from the coupling to that state \cite{jidocal} and we settle for the effective coupling
\begin{equation*}
  \tilde{g}_{\Lambda^* \pi \Sigma}=-1.54-0.93i.
\end{equation*}

\subsection{Other terms in the $\gamma p \rightarrow K^+\Lambda(1405)$ amplitude}

\begin{figure}[t]
\includegraphics[width=0.3\textwidth]{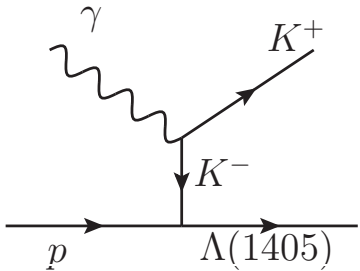}
\caption{Diagram of $K$-exchange.}
\label{fig:kexchange}
\end{figure}

We do not want to make an elaborate model for the process, but will introduce three terms, corresponding to physical mechanisms, but with some flexibility  in the parameters, such that one has a general structure for what we call ``background" terms in the region of the peak. The first one is the $K$-exchange depicted in Fig.~\ref{fig:kexchange}.

Using the same Lagrangians described in prior sections, we find,
\begin{eqnarray}
-i t_{\gamma K^+K^-}=ie(p_{K^+}-p_{K^-})_j \,\epsilon_j(\gamma)
\end{eqnarray}
and in the Coulomb gauge which we use to comply with gauge invariance, we find,
\begin{equation}
-it_K = ie\,2g_{\Lambda^*K^-p}\frac{1}{(p_\gamma-k)^2-m_K^2}\vec{\epsilon}(\gamma)\cdot \vec{k}.
\end{equation}

Once again we follow the same procedure as before, to take into account the two $\Lambda(1405)$ states. Since the $\Lambda^*$ is observed in $\pi\Sigma$, now the relevant amplitude is $K^-p\to\pi\Sigma$, which is dominated by the second pole around 1420~MeV. Now we have,
\begin{eqnarray}
t_{\bar{K}N,\pi\Sigma}&=&\frac{\tilde{g}_{\Lambda^*\bar{K}N} \tilde{g}_{\Lambda^*\pi\Sigma}}{\sqrt{s}-\tilde{M}_{\Lambda^*}+i\frac{\tilde{\Gamma}_{\Lambda^*}}{2}},
\end{eqnarray}
and using the results of Ref.~\cite{luis1}, we get,
\begin{eqnarray}
\tilde{\Gamma}_{\Lambda^*}=40~{\rm~MeV}, ~~~~~|\tilde{g}_{\Lambda^*\bar{K}N}|^2=3.1.
\end{eqnarray}
By giving this coupling the phase of the coupling of the second pole, and taking the isospin Clebsch-Gordan coefficient for the $K^-p$ component, we get,
\begin{equation}
\tilde{g}_{\Lambda^*K^-p} \approx \frac{1}{\sqrt{2}}\left(-1.65+0.62i\right).
\end{equation}

Another mechanism which was used in Ref.~\cite{nacher} was the contact term reflected in the diagrams of Fig.~5, which was used at lower photon energies.
\begin{figure}[t]
\includegraphics[width=0.45\textwidth]{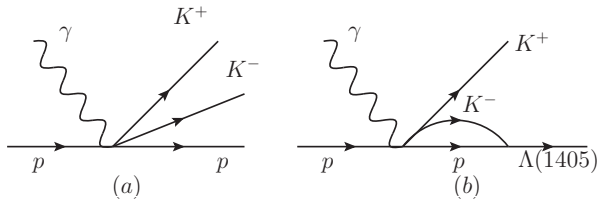}
\caption{(a) Contact term for the $\gamma p \to K^+ K^- p$ reaction, (b) Diagram leading to the formation of the $K^+\Lambda^*$ in the final state.}
\label{fig:other}
\end{figure}

The contact term for the diagram of Fig.~5(a) is given by~\cite{nacher},
\begin{equation}
t=-2i\frac{(\vec{\sigma}\times \vec{p}_\gamma)\cdot \vec{\epsilon}(\gamma)}{2M_p}\,\frac{e}{4f^2_\pi},
\end{equation}
and for the diagram of Fig.~5(b), by
\begin{equation}
t_c=-2i \frac{(\vec{\sigma}\times \vec{p}_\gamma)\cdot \vec{\epsilon}(\gamma)}{2M_p}\,\frac{e}{4f^2_\pi} G_{K^-p} \tilde{g}_{\Lambda^*K^-p},
\end{equation}
where $G_{K^-p}$ is the loop function for the intermediate $K^-p$ state~\cite{angels}. The intermediate states with neutral mesons do not contribute to this mechanism and there is cancellation between the channels $\pi^-\Sigma^+$ and $\pi^+\Sigma^-$.

\begin{figure}[t]
\includegraphics[width=0.3\textwidth]{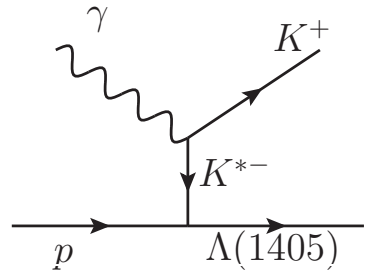}
\caption{Diagram for $K^*$-exchange.}
\label{fig:kstar}
\end{figure}

Finally we also consider $K^*$-exchange as depicted in Fig.~\ref{fig:kstar}.
In this case, we do not have a theoretical coupling for $K^*p\to \Lambda^*$, but the structure is of the $\vec{\sigma}\cdot \vec{\epsilon}\,(K^*)$ type. The $\gamma K^* K$ vertex is of anomalous  type involving a Lagrangian of the type

\begin{equation}
\epsilon_{\mu\nu\alpha\beta}  \partial^\mu \epsilon^\nu (\gamma)\partial^\alpha \epsilon^\beta(K^*).
\end{equation}

Altogether we get a term of the type
\begin{equation}
t_{K^*} =  (\vec{\sigma} \times\vec{p}_\gamma)\cdot\vec{\epsilon}(\gamma) \frac{1}{(p_\gamma-k)^2-m^2_{K^*}}.
\end{equation}
The propagators in the $K$- and $K^*$-exchanges peak at forward angle.

We note that,
\begin{equation}
\overline{\sum}\sum \left[\left(\vec{\sigma}\times \vec{p}_\gamma \right) \cdot \vec{\epsilon}(\gamma) \right]^2= \vec{p}^{\,2}_\gamma
\end{equation}
and there is no interference between the $\vec{\sigma}$ terms and the non $\vec{\sigma}$ terms.

We shall conduct a fit to the data, ${\rm d}\sigma/{\rm dcos}\theta$, multiplying the different terms by a factor to incorporate other terms not accounted for explicitly with a similar structure. Thus we write for the total amplitude,
\begin{equation}
t=at_T + b t_K + ct_c + dt_{K^*},
\label{eq:fullamp}
\end{equation}
where $a$, $b$, $c$ and $d$ will be free parameters.

We have seen that in the contribution of $t_T$ for the triangle singularity, there are many ingredients, several couplings, approximations done, such that we allow the fit to choose a value of $a \neq 1$, but not too different and so can we say the same about  the $b$ coefficient.

Finally, in some option, we shall also introduce a form factor for $K$- and $K^*$-exchanges,
\begin{equation}
F(q)=\frac{\Lambda^2}{\Lambda^2+\vec{q}^{\,2}}
\end{equation}
with $\vec q$ the momentum of the $K$ or  $K^*$ boosted to the rest frame of the $\Lambda^*$.

\section{Results}
\begin{figure}[t]
\includegraphics[width=0.45\textwidth]{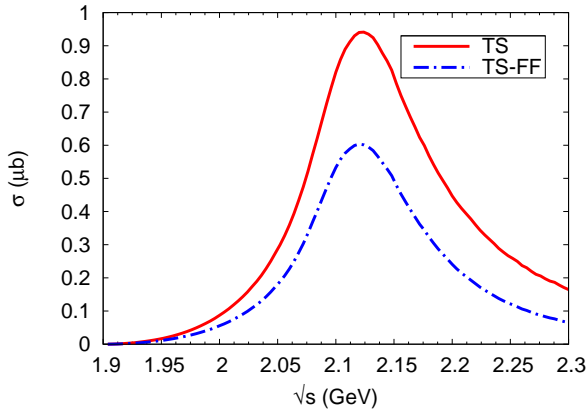}
\caption{Results for $\sigma$ obtained with the triangle diagram amplitude $t_T$. Solid line for $t_T$ of the text. Dashed line: two form factors of the type $\Lambda^2/(\Lambda^2+\vec{q}^{\,2})$ have been implemented in addition, with $\Lambda=0.8$ GeV.}
\label{fig:tcsTS}
\end{figure}

In the first place we calculate the integrated cross section with just the triangle mechanism, the amplitude $t_T$ of Eq.~(\ref{eq:t_T3}).  The results are shown in Fig.~\ref{fig:tcsTS}. The two curves correspond to $t_T$ of the text (solid line) and the same adding a form factor $\Lambda^2/(\Lambda^2+\vec{q}^{\, 2})$ for each of the vertices $N^*\to K^*\Sigma$ and $K^*\to K \pi$. It is interesting to observe that the shape is reasonably similar to the experimental one (see Fig.~16 of Ref.~\cite{moriya}) around $\sqrt{s}=2110$~MeV and the strength also very similar to the experimental one (0.6 $\mu$b at the peak). Of course we know that not all of the strength comes from this mechanism, but the results obtained indicate that this is an important contribution not to be missed. As indicated in the preceding section, the many couplings involved and some approximation done to evaluate the mechanism  give us reason to accept that moderate changes in the strength should be in order. These changes will be accommodated by changing the coefficient $a$ of Eq.~(\ref{eq:fullamp}) in the fit to the data. The coefficient $a$ will be applied to the result with $t_T$ of the text without form factor.

\begin{figure}[t]
\includegraphics[width=0.45\textwidth]{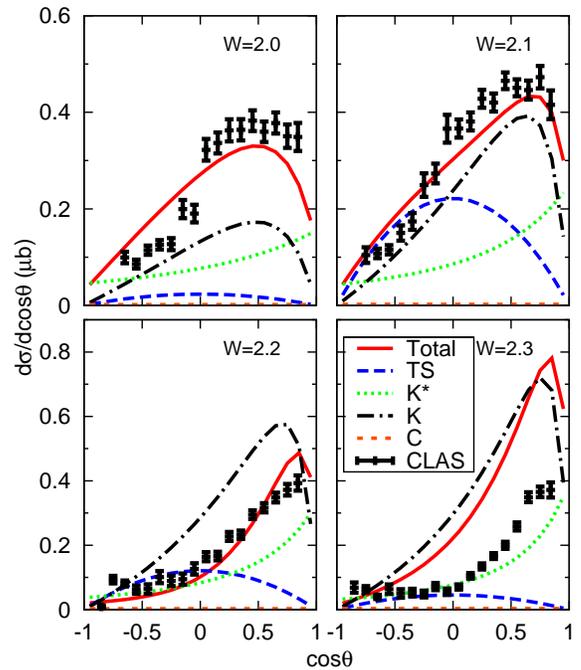}
\caption{The results of the fit to the differential cross section data ($\sqrt{s}=2000,2100,2200$~MeV~\cite{moriya}) for the $\gamma p\to K^+\Lambda(1405)$ reaction. TS, triangle singularity; $K^*$, $K^*$ exchange; $K$, $K$ exchange; C, contact term; solid line, total.}
\label{fig:dcs}
\end{figure}

\begin{figure}[t]
\includegraphics[width=0.45\textwidth]{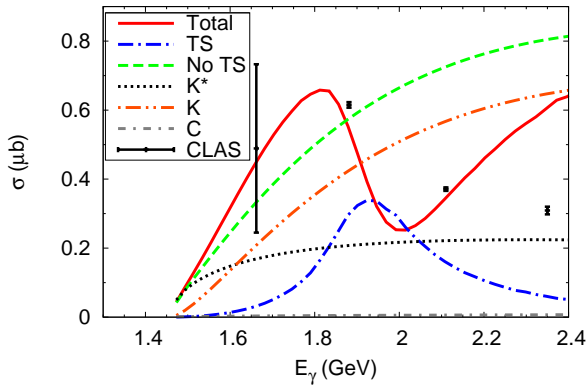}
\caption{The predicted integral cross section for the $\gamma p\to K^+\Lambda(1405)$ reaction.}
\label{fig:tcs}
\end{figure}

\begin{figure}[t]
\includegraphics[width=0.45\textwidth]{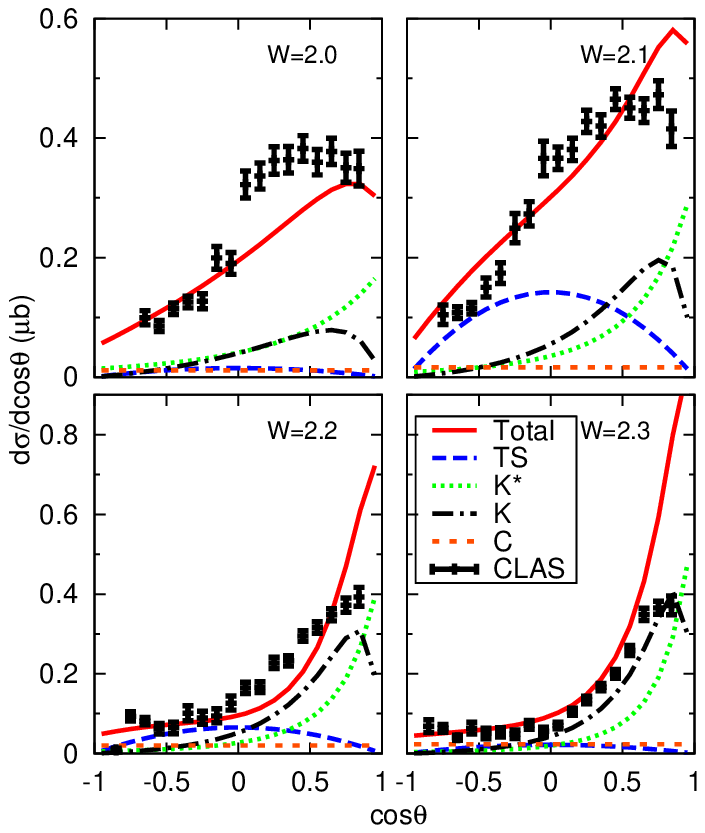}
\caption{Same as Fig.~\ref{fig:dcs}, but with the form factor in $K$- and $K^*$-exchanges.}
\label{fig:dcs_FF}
\end{figure}

\begin{figure}[t]
\includegraphics[width=0.4\textwidth]{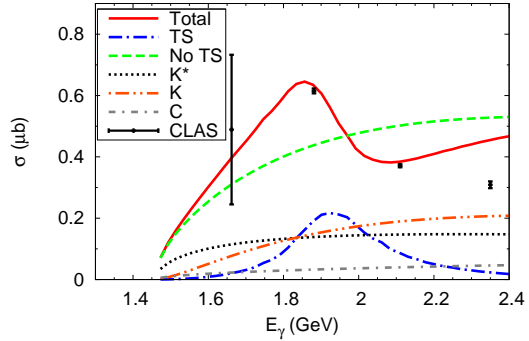}
\caption{Same as Fig.~\ref{fig:tcs}, but with the form factor in $K$- and $K^*$-exchanges.}
\label{fig:tcs_FF}
\end{figure}

In Fig.~\ref{fig:dcs}, we show a fit to ${\rm d}\sigma/{\rm dcos}\theta$ for three energies around $\sqrt{s}=2100$~MeV, which are 2000, 2100, and 2200~MeV. The experimental data are averaged over a span of 100~MeV around the centroid  and we do the same. We also show the results for $\sqrt{s}=2300$~MeV, but these data are outside the range of interest to us and we do not include them in the fit. The agreement with the data obtained is rather good (for the energies fitted), when one considers that there are appreciable differences between ${\rm d}\sigma/{\rm dcos}\theta$ for $\pi^+\Sigma^-$, $\pi^-\Sigma^+$, $\pi^0\Sigma^0$ and all of them are summed up in the data of the figures. Our results for the data of the last energy, which have not been fitted, are bigger than those of  experiment but of the right order.

What is relevant for us is that the fit returns the coefficients $a=0.6$, $b=1.5-0.8i$, $c=0.3$, $d=11.7$~\footnote{
Note that the $a$ and $b$ terms of Eq.~(\ref{eq:fullamp}) have the same structure, and do not interfere with the $c$ and $d$ terms. It suffices for generality to take one of the two coefficients complex, and we have chosen $b$. The same can be said about $c$ and $d$, and we take $d$ real,  for the largest term of the two. The term with $c$ has a small strength and the fit is compatible with $c$ real. 
}. 
This means that the fit requires an acceptable fraction of the triangle singularity contribution. It is also interesting to see that the  triangle singularity contribution is small for the $\sqrt{s}=2000$~MeV band, quite large for the $\sqrt{s}=2100$~MeV band, as expected, and smaller in the $\sqrt{s}=2200$~MeV band. For $\sqrt{s}=2300$~MeV, it becomes again very small. The contribution of the contact term is small and we do not discuss it further. One can see that the $K$-exchange term gives a large contribution and is responsible for  the large strength at $\sqrt{s}=2300$~MeV. Should we implement form factors in the mechanism, its contribution would be smaller and we do that in a new step. It is also interesting to see that the contribution of the $K^*$-exchange has become moderate. In Fig.~\ref{fig:tcs}, we also show the integrated cross section over angle and compare it with the data. We see that the agreement is fair for the three energies where we fitted ${\rm d}\sigma/{\rm dcos}\theta$.

We should stress that the agreement found for ${\rm d}\sigma/{\rm dcos}\theta$ is not trivial since there are interferences between the $K$-exchange and the triangle diagram. At energies $\sqrt{s}=2300$~MeV
 and beyond, the cross sections with this simplified model grow up but stabilize around 0.8 $\mu$b. However this is a regime that we are not interested in, which definitely would need improvements, but what matters for our purpose is that the contribution of the triangle singularity becomes negligible.

In Fig.~\ref{fig:dcs_FF}, we conduct another fit, introducing now the form factor in the $K$- and $K^*$-exchanges. The fit gives us now $a=0.6$, $b=1.9+0.2i$, $c=-0.8$ and $d=17.6$. The contribution of the $K$-exchange is now smaller and is compensated by a larger contribution of $K^*$, still moderate.

The angle integrated cross section in this latter case is shown in Fig.~\ref{fig:tcs_FF}. 
The cross section of the $\sqrt{s}=2300$~MeV band is now better than before, and for higher energies, the cross section stabilizes around 0.6 $\mu$b. Yet, the most important point is that the strength of the triangle singularity needed in the fit is the same as before, with the coefficient $a=0.6$.

We have conducted other fits, one of them includes fitting the data in the band of $\sqrt{s}=2300$~MeV. One gets a better agreement at higher energies at the expense of a somewhat worse agreement in the band of $\sqrt{s}=2000$~MeV. Another fit is done assuming a smaller width for the $N^*(2030)$, of the order of 200~MeV. The important outcome from all these fits is that the strength of the triangle singularity needed remains the same. All these results confirm the relevance of the triangle mechanism with a strength compatible with the calculated one within the estimated uncertainties.

One could take more background terms to have a more complete model. However, before doing that, it is instructive to see what has been done in different models in the literature on the subject. In Ref.~\cite{nam}, the authors take $K$- and $K^*$-exchanges, as in the present case, but in addition, they have an $s$-channel term with the nucleon pole, and a $u$-term with the $\Lambda^*$ pole. The model produces a cross section that raises from threshold and falls down monotonically  beyond $E_\gamma=1.6$~GeV. In the region around $E_\gamma=1.8\sim 1.9$~GeV,  the cross section is smooth, independently of the  choice of parameters that they use, and the cross section is limited between $0.1 \sim 0.2~\mu$b, short of the experimental results shown in Fig.~\ref{fig:tcs}. The works of Refs.~\cite{luis1,luis2,mai} do not use an explicit amplitude and parametrize the strength of the $\gamma N \to$ meson + baryon vertex, prior to the final state interaction of the meson-baryon pairs that generates the $\Lambda(1405)$. The strength of these primary vertices is chosen for each energy and hence the origin of the peak in the cross section cannot be traced down with this approach. In the most detailed model of Ref.~\cite{jido}, other terms are taken, apart from $K$- and $K^*$-exchanges, and several contact terms are introduced by hand which are fitted for each energy. Once again with this strategy, one cannot assess the origin of the peak in the cross section. Even then, it is clarifying  the fact that with all this freedom still the angular dependence in the region of the peak could not be reproduced. What we have done here is to evaluate a new and unavoidable contribution stemming from a triangle singularity. Within the accepted small uncertainties in its strength, we see that it plays an important role in providing both, the peak in the cross section and the angular dependence, something that other models incorporating more terms in the amplitudes than we have considered, fail to reproduce. 

\section{Conclusions}
We have performed a study of the contribution of a triangle diagram to the $\gamma p\to K^+\Lambda(1405)$ reaction. The mechanism consists of the formation of the resonance $N^*(2030)$, predicted theoretically within the local hidden gauge approach to vector-baryon interaction, and supported by results of $\gamma p\to K^0\Sigma^+$ done at the  $K^*\Lambda$ and $K^*\Sigma$ thresholds. This resonance couples strongly to $K^*\Sigma$ and produces a singularity via the triangle mechanism where the $N^*(2030)$ decays into $K^*\Sigma$, the $K^*$ decays to $K^+\pi$ and the $\pi\Sigma$ merge to form the $\Lambda(1405)$. The peak of the singularity shows up around $\sqrt{s}=2110$~MeV and the mechanism also has its largest strength around ${\rm cos}\theta=0$. This is precisely the region of energies and the angles where conventional models failed to reproduce the experimental data. We have shown that adding a few basic mechanisms to the triangle one, and with a strength for this latter mechanism compatible with theoretical uncertainties, we find a good agreement for ${\rm d}\sigma/{\rm dcos}\theta$ and the integrated cross section around the experimental peak at $\sqrt{s}=2110$~MeV. 
By looking into different models in the literature, we see that explicit models incorporating more background terms than we have considered here fail to produce the peak in the cross section. We also trace back the apparent good agreement of other models with the fact that free parameters are adjusted for each energy, and even then they fail to provide a good angular dependence in the region of the peak. The mechanism described here provides a microscopical explanation of both, the peak around $E_\gamma=1.8$~GeV, and the angular dependence around this energy region.

\section*{Acknowledgments}
We would like to thank R. Schumacher for useful comments and for pointing us some details on the data.
One of us, E. O. wishes to acknowledge support from the Chinese Academy
of Science in the Program of Visiting Professorship for Senior International Scientists (Grant No. 2013T2J0012).
This work is partly supported by the
National Natural Science Foundation of China under Grants
Nos. 11565007, 11547307, 11475227 and 11505158 . It is also supported by the Youth Innovation Promotion Association CAS (No. 2016367),
by DFG and NSFC through funds provided to the
Sino-German CRC 110 ``Symmetries and the Emergence of Structure in QCD'' (NSFC
Grant No.~11621131001), by the Chinese Academy of Sciences (Grant
No.~QYZDB-SSW-SYS013), by the Thousand
Talents Plan for Young Professionals,
by the China Postdoctoral Science Foundation (No. 2015M582197) and the Postdoctoral Research Sponsorship in Henan Province (No. 2015023).
This work is also partly supported by the Spanish Ministerio
de Economia y Competitividad and European FEDER funds
under the contract number FIS2011-28853-C02-01, FIS2011-
28853-C02-02, FIS2014-57026-REDT, FIS2014-51948-C2-
1-P, and FIS2014-51948-C2-2-P, and the Generalitat Valenciana
in the program Prometeo II-2014/068.


\begin{thebibliography}{99}

\bibitem{Landau:1959fi}
  L.~D.~Landau,
  Nucl.\ Phys.\  {\bf 13}, 181 (1959).



\bibitem{Coleman:1965xm}
  S.~Coleman and R.~E.~Norton,
  Nuovo Cim.\  {\bf 38}, 438 (1965).

\bibitem{guobayar} 
  M.~Bayar, F.~Aceti, F.~K.~Guo and E.~Oset,
  Phys.\ Rev.\ D {\bf 94}, no. 7, 074039 (2016).
\bibitem{BESIII:2012aa}
  M.~Ablikim {\it et al.} [BESIII Collaboration],
  Phys.\ Rev.\ Lett.\  {\bf 108}, 182001 (2012).


\bibitem{Wu:2011yx}
  J.~J.~Wu, X.~H.~Liu, Q.~Zhao and B.~S.~Zou,
  Phys.\ Rev.\ Lett.\  {\bf 108}, 081803 (2012).


\bibitem{Wu:2012pg}
  X.~G.~Wu, J.~J.~Wu, Q.~Zhao and B.~S.~Zou,
  Phys.\ Rev.\ D {\bf 87},  014023 (2013).


\bibitem{Aceti:2012dj}
  F.~Aceti, W.~H.~Liang, E.~Oset, J.~J.~Wu and B.~S.~Zou,
  Phys.\ Rev.\ D {\bf 86}, 114007 (2012).

\bibitem{Liu:2015taa}
  X.~H.~Liu, M.~Oka and Q.~Zhao,
  Phys.\ Lett.\ B {\bf 753}, 297 (2016).

\bibitem{Ketzer:2015tqa}
  M.~Mikhasenko, B.~Ketzer and A.~Sarantsev,
  Phys.\ Rev.\ D {\bf 91},  094015 (2015).


\bibitem{Aceti:2016yeb} 
  F.~Aceti, L.~R.~Dai and E.~Oset,
  Phys.\ Rev.\ D {\bf 94}, no. 9, 096015 (2016).


\bibitem{Adolph:2015pws}
  C.~Adolph {\it et al.} [COMPASS Collaboration],
  Phys.\ Rev.\ Lett.\  {\bf 115},  082001 (2015).

\bibitem{Aaij:2015tga}
  R.~Aaij {\it et al.} [LHCb Collaboration],
  Phys.\ Rev.\ Lett.\  {\bf 115}, 072001 (2015).


\bibitem{Aaij:2015fea}
  R.~Aaij {\it et al.} [LHCb Collaboration],
  Chin.\ Phys.\ C {\bf 40},  011001 (2016).


\bibitem{Guo:2015umn}
  F.-K.~Guo, U.-G.~Mei{\ss}ner, W.~Wang and Z.~Yang,
  Phys.\ Rev.\ D {\bf 92},  071502 (2015).


\bibitem{Liu:2015fea}
  X.~H.~Liu, Q.~Wang and Q.~Zhao,
  Phys.\ Lett.\ B {\bf 757}, 231 (2016).


\bibitem{Guo:2016bkl} 
  F.~K.~Guo, U.-G.~Mei{\ss}ner, J.~Nieves and Z.~Yang,
  Eur.\ Phys.\ J.\ A {\bf 52}, no. 10, 318 (2016).

\bibitem{Aaij:2016ymb}
  R.~Aaij {\it et al.} [LHCb Collaboration],
  Phys.\ Rev.\ Lett.\  {\bf 117}, 082003 (2016).


\bibitem{Aaij:2014zoa}
  R.~Aaij {\it et al.} [LHCb Collaboration],
  JHEP {\bf 1407}, 103 (2014).

\bibitem{moriya}
  K.~Moriya {\it et al.} [CLAS Collaboration],
  Phys.\ Rev.\ C {\bf 88}, 045201 (2013)
  Addendum: [Phys.\ Rev.\ C {\bf 88}, 049902 (2013)].

\bibitem{nacher}
  J.~C.~Nacher, E.~Oset, H.~Toki and A.~Ramos,
  Phys.\ Lett.\ B {\bf 455}, 55 (1999).

\bibitem{niiyama}
  M.~Niiyama {\it et al.},
  Phys.\ Rev.\ C {\bf 78}, 035202 (2008).

\bibitem{Schumacher:2013vma}
  R.~A.~Schumacher and K.~Moriya,
  Nucl.\ Phys.\ A {\bf 914}, 51 (2013).

\bibitem{luis1}
  L.~Roca and E.~Oset,
  Phys.\ Rev.\ C {\bf 87},  055201 (2013).

\bibitem{luis2}
  L.~Roca and E.~Oset,
  Phys.\ Rev.\ C {\bf 88},  055206 (2013).

\bibitem{mai}
  M.~Mai and U.-G.~Mei{\ss}ner,
  Eur.\ Phys.\ J.\ A {\bf 51},  30 (2015).

\bibitem{ollerulf}
  J.~A.~Oller and U.-G.~Mei{\ss}ner,
  Phys.\ Lett.\ B {\bf 500}, 263 (2001).

\bibitem{jidocal}
  D.~Jido, J.~A.~Oller, E.~Oset, A.~Ramos and U.-G.~Mei{\ss}ner,
  Nucl.\ Phys.\ A {\bf 725}, 181 (2003).

\bibitem{carmen}
  C.~Garcia-Recio, J.~Nieves, E.~Ruiz Arriola and M.~J.~Vicente Vacas,
  Phys.\ Rev.\ D {\bf 67}, 076009 (2003).



\bibitem{hyodo}
  T.~Hyodo, S.~I.~Nam, D.~Jido and A.~Hosaka,
  Phys.\ Rev.\ C {\bf 68}, 018201 (2003).

\bibitem{borawolfrom}
  B.~Borasoy, R.~Nissler and W.~Weise,
  Eur.\ Phys.\ J.\ A {\bf 25}, 79 (2005).



\bibitem{boraulf}
  B.~Borasoy, U.-G.~Mei{\ss}ner and R.~Nissler,
  Phys.\ Rev.\ C {\bf 74}, 055201 (2006).

\bibitem{ikada}
  Y.~Ikeda, T.~Hyodo and W.~Weise,
  Nucl.\ Phys.\ A {\bf 881}, 98 (2012).



\bibitem{cotanch}
  R.~A.~Williams, C.~R.~Ji and S.~R.~Cotanch,
  Phys.\ Rev.\ C {\bf 43}, 452 (1991).

\bibitem{choi}
  T.~K.~Choi, K.~S.~Kim and B.~G.~Yu,
  arXiv:0911.0083 [nucl-th]. 


\bibitem{nam}
  S.~i.~Nam, J.~H.~Park, A.~Hosaka and H.~C.~Kim,
  J.\ Korean Phys.\ Soc.\  {\bf 59}, 2676 (2011).
  
\bibitem{jido}
  S.~X.~Nakamura and D.~Jido,
  PTEP {\bf 2014}, 023D01 (2014).

\bibitem{tokinacher}
  J.~C.~Nacher, E.~Oset, H.~Toki and A.~Ramos,
  Phys.\ Lett.\ B {\bf 461}, 299 (1999).

\bibitem{borasoy}
  B.~Borasoy, P.~C.~Bruns, U.-G.~Mei{\ss}ner and R.~Nissler,
  Eur.\ Phys.\ J.\ A {\bf 34}, 161 (2007).

\bibitem{mori2}
  K.~Moriya {\it et al.} [CLAS Collaboration],
  Phys.\ Rev.\ C {\bf 87},  035206 (2013).

\bibitem{angelsvec}
  E.~Oset and A.~Ramos,
  Eur.\ Phys.\ J.\ A {\bf 44}, 445 (2010).

\bibitem{angelselsa}
  A.~Ramos and E.~Oset,
  Phys.\ Lett.\ B {\bf 727}, 287 (2013).

\bibitem{ewald}
  R.~Ewald {\it et al.} [CBELSA/TAPS Collaboration],
  Phys.\ Lett.\ B {\bf 713}, 180 (2012).



\bibitem{garzon}
  E.~J.~Garzon and E.~Oset,
  Eur.\ Phys.\ J.\ A {\bf 48}, 5 (2012).

\bibitem{lianguchino}
  T.~Uchino, W.~H.~Liang and E.~Oset,
  Eur.\ Phys.\ J.\ A {\bf 52},  43 (2016).


\bibitem{amsler}
  C.~Amsler {\it et al.} [Particle Data Group],
  Phys.\ Lett.\ B {\bf 667}, 1 (2008).


\bibitem{hidden1}
  M.~Bando, T.~Kugo, S.~Uehara, K.~Yamawaki and T.~Yanagida,
  Phys.\ Rev.\ Lett.\  {\bf 54}, 1215 (1985).

\bibitem{hidden2}
  M.~Bando, T.~Kugo and K.~Yamawaki,
  Phys.\ Rept.\  {\bf 164}, 217 (1988).

\bibitem{hidden4}
  U.-G.~Mei{\ss}ner,
  Phys.\ Rept.\  {\bf 161}, 213 (1988).

\bibitem{hideko}
  H.~Nagahiro, L.~Roca, A.~Hosaka and E.~Oset,
  Phys.\ Rev.\ D {\bf 79}, 014015 (2009).

\bibitem{sakurai}
J.J. Sakurai, Currents and Mesons, University of Chicago Press, 1969.

\bibitem{fcajorgi}
  F.~Aceti, J.~M.~Dias and E.~Oset,
  Eur.\ Phys.\ J.\ A {\bf 51},  48 (2015).

\bibitem{angels}
  E.~Oset and A.~Ramos,
  Nucl.\ Phys.\ A {\bf 635}, 99 (1998).

\bibitem{hyodoulf}
U.-G, Mei{\ss}ner, T. Hyodo, {\it Pole Structure of the $\Lambda(1405)$ Region}, in 
  C.~Patrignani, {\it et al.} [Particle Data Group]
  Chin.\ Phys.\ C {\bf 40},  100001 (2016).



\end{thebibliography}
\end{document}